\documentclass{article}
\usepackage{amsmath, amsthm}

\usepackage{graphicx}
\newtheoremstyle{theorem}
  {10pt}		  
  {10pt}  
  {\sl}  
  {\parindent}     
  {\bf}  
  {. }    
  { }    
  {}     
\theoremstyle{theorem}

\newtheoremstyle{defi}
  {10pt}		  
  {10pt}  
  {\rm}  
  {\parindent}     
  {\bf}  
  {. }    
  { }    
  {}     
\theoremstyle{defi}

\begin{document}

\title{Viable models of traversable wormholes supported by small amounts of exotic matter}

\author{Peter K.F. Kuhfittig\\
Department of Mathematics\\
Milwaukee School of Engineering\\
Milwaukee, WI  53202-3109 USA\\
kuhfitti@msoe.edu\\[2pt]}
\date{}

\maketitle
\begin{abstract}\noindent
Wormholes allowed by the general theory of relativity that are 
simultaneously traversable by humanoid travelers are subject to severe 
constraints from quantum field theory, particularly the so-called 
quantum inequalities, here slightly extended.  Moreover, self-collapse 
of such wormholes can only be prevented by the use of ``exotic matter," 
which, being rather problematical, should be used in only minimal 
quantities.  However, making the layer of exotic matter arbitrarily 
thin leads to other problems, such as the need for extreme 
fine-tuning.  This paper discusses a class of wormhole geometries 
that strike a balance between reducing the proper distance across 
the exotic region and the degree of fine-tuning required to achieve 
this reduction. Surprisingly, the degree of fine-tuning appears to 
be a generic feature of the type of wormhole discussed.  No particular 
restriction is placed on the throat size, even though the proper 
thickness of the exotic region can indeed be quite small.  Various 
traversability criteria are shown to be met.


\end{abstract}

\phantom{a}
PAC numbers: 04.20.Jb, 04.20.Gz
\section{Introduction: viable wormhole models at last}
It is well known that traversable wormholes require exotic matter to 
prevent self-collapse \cite{MT88}.  Such matter is confined to a small 
region around the throat, a region in which the weak energy condition 
is violated.  While it is desirable to keep this region as small as 
possible, the use of arbitrarily small amounts of exotic matter leads 
to severe problems, as discussed by Fewster and Roman 
\cite{FR05a, FR05b}.  The discovery by Ford and Roman \cite{FR95,FR96}
that quantum field theory places severe constraints on the wormhole 
geometries has shown that most of the ``classical" wormholes could not 
exist on a macroscopic scale.  The wormhole described by Kuhfittig 
\cite{pK06} is an earlier attempt to strike a balance between two 
conflicting requirements, reducing the amount of exotic matter and 
fine-tuning the values of certain parameters.  The purpose of this 
paper is to extend these ideas to much more general models.  The
quantum inequalities are generalized to be valid, not only at the 
throat, but in the entire exotic region.  The models discussed will 
therefore (1) satisfy all the constraints imposed by quantum field 
theory, (2) strike a reasonable balance between a small proper 
thickness of the exotic region and the degree of fine-tuning of the 
metric coefficients, Eq.~(\ref{E:line}) below, and (3) minimize the 
assumptions on these metric coefficients.  Another key finding is 
that the degree of fine-tuning is the same for all of the wormholes 
models considered.

Problems with arbitrarily small amounts of exotic matter are also 
discussed in Ref. \cite{oZ07}, but the author states explicitly 
that the issues discussed here and in Ref. \cite{pK06} are beyond 
the scope of his paper.  
 
\section{A general model}
Consider the general line element \cite{pK02}
\begin{equation}\label{E:line}
   ds^2 =-e^{2\gamma(r)}dt^2+e^{2\alpha(r)}dr^2+r^2(d\theta^2+
      \text{sin}^2\theta\, d\phi^2),
\end{equation}
where the units are taken to be those for which $G=c=1.$  The function 
$\gamma$ is called the redshift function; this function must be 
everywhere finite to avoid an event horizon at the throat.  The function 
$\alpha$ is related to the shape function $b=b(r)$: 
\[
   e^{2\alpha(r)}=\frac{1}{1-b(r)/r}.
\]
(The shape function determines the spatial shape of the wormhole when 
viewed, for example, in an embedding diagram.)  It now follows that 
\begin{equation}\label{E:shape}
    b(r)=r(1-e^{-2\alpha(r)})
\end{equation}
and that $\alpha$ has a vertical asymptote at the throat $r=r_0$:
\[ 
   \lim_{r \to r_0+}\alpha(r)=+\infty.
\]
Also, $\alpha(r)\rightarrow 0$ and $\gamma(r)\rightarrow 0$ as 
$r\rightarrow \infty$.  The qualitative features of $\alpha(r)$, 
$\gamma(r)$, and $-\gamma(r)$, the reflection of $\gamma(r)$ in 
the horizontal axis, are shown in Fig.~1.  It is assumed that
$\alpha$ and $\gamma$ are twice differentiable with $\alpha'(r)<0$ 
and $\gamma'(r)>0$; in addition, $\alpha''(r)>0$, 
$\gamma''(r)\le 0$, and $\alpha''(r)>|\gamma''(r)|$.
\begin{figure}[htbp]
\begin{center}
\includegraphics[clip=true, draft=false, bb=0 0 299 212, angle=0, width=4.24in, height=3in, 
   viewport=50 50 296 200]{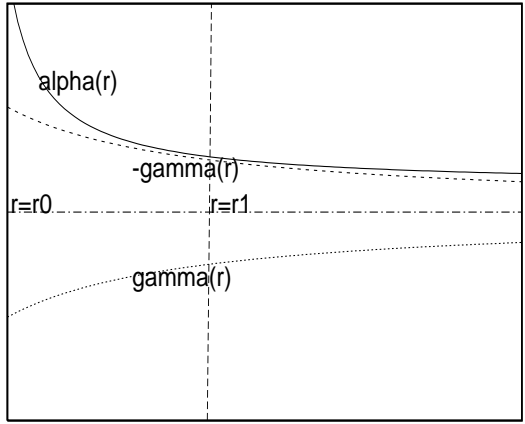}
\end{center}
\caption{\label{fig:figure1}Qualitative features of $\alpha(r)$ and $\gamma(r)$.}
\end{figure} 
  
 The next step is to list the components of the Einstein tensor in the 
orthonormal frame.  From Ref.~\cite{pK02},
\[
   G_{\hat{t}\hat{t}}=\frac{2}{r}e^{-2\alpha(r)}\alpha'(r)
   +\frac{1}{r^2}(1-e^{-2\alpha(r)}),
\]
\[
    G_{\hat{r}\hat{r}}=\frac{2}{r}e^{-2\alpha(r)}\gamma'(r)
   -\frac{1}{r^2}(1-e^{-2\alpha(r)}),
\]
and
\begin{equation*}
   G_{\hat{\theta}\hat{\theta}}=G_{\hat{\phi}\hat{\phi}}=
    e^{-2\alpha(r)}\left[\gamma''(r)
      +\alpha'(r)\gamma'(r)\phantom{\frac{1}{r}}\right.\\
     \left.+[\gamma'(r)]^2+\frac{1}{r}\gamma'(r)
      -\frac{1}{r}\alpha'(r)\right].   
\end{equation*}
Since the Einstein field equations $G_{\hat{\alpha}\hat{\beta}}=
8\pi T_{\hat{\alpha}\hat{\beta}}$ imply that the stress-energy tensor 
is proportional to the Einstein tensor, the only nonzero components are 
$T_{\hat{t}\hat{t}}=\rho,$ $T_{\hat{r}\hat{r}}=-\tau,$ and 
$T_{\hat{\theta}\hat{\theta}}=T_{\hat{\phi}\hat{\phi}}=p.$
Now recall that the weak energy condition (WEC) requires the 
mass-energy tensor $T_{\alpha\beta}$ to obey 
\[
     T_{\alpha\beta}\mu^{\alpha}\mu^{\beta}\ge0  
\]
for all time-like vectors and, by continuity, all null vectors. 
Using the radial outgoing null vector $\mu^{\hat{\alpha}}=(1,1,0,0)$, 
the condition now becomes $T_{\hat{t}\hat{t}}+T_{\hat{r}\hat{r}}=
\rho-\tau\ge 0.$  So if the WEC is violated, then $\rho-\tau<0$.
The field equations $G_{\hat{\alpha}\hat{\beta}}=
8\pi T_{\hat{\alpha}\hat{\beta}}$ now imply that 
\begin{equation}\label{E:WEC}
  \rho-\tau=\frac{1}{8\pi}\left[\frac{2}{r}e^{-2\alpha(r)}
    \left[\alpha'(r)+\gamma'(r)\right]\right].
\end{equation}
Sufficiently close to the asymptote, $\alpha'(r)+\gamma'(r)$ is 
clearly negative.  (Recall that $\alpha'<0$ and $\gamma'>0$.)
To satisfy the Ford-Roman constraints \cite{FR95, FR96}, we would 
like the WEC to be satisfied outside of some small interval $[r_0,r_1]$.  
In other words, 
\begin{equation}\label{E:FR1}
    |\alpha'(r_1)|=\gamma'(r_1),
\end{equation}
\begin{equation}\label{E;FR2}
    \alpha'(r)+\gamma'(r)<0\quad \text{for}\quad r_0<r<r_1,
\end{equation}
and
\begin{equation}\label{E:FR3}
    \alpha'(r)+\gamma'(r)>0 \quad \text{for} \quad r>r_1.
\end{equation}
(See Fig.~1.) 
\section{Other constraints}\label{S:other}
Before discussing additional constraints, we need to list some of 
the  components of the Riemann curvature tensor in the orthonormal
frame.  From Ref.~\cite{pK02}
\begin{equation}\label{E:Riemann1}
  R_{\hat{r}\hat{t}\hat{r}\hat{t}}=e^{-2\alpha(r)}
   \left(\gamma''(r)-\alpha'(r)\gamma'(r)
      +\left[\gamma'(r)\right]^2\right),
\end{equation}
\begin{equation}\label{E:Riemann2}
   R_{\hat{\theta}\hat{t}\hat{\theta}\hat{t}}=\frac{1}{r}
     e^{-2\alpha(r)}\gamma'(r),
\end{equation}
and
\begin{equation}\label{E:Riemann3}
   R_{\hat{\theta}\hat{r}\hat{\theta}\hat{r}}=\frac{1}{r}
       e^{-2\alpha(r)}\alpha'(r).      
\end{equation}
Much of what follows is based on the discussion in 
Ref. \cite{MT88}.  In particular, we have for the radial tidal 
constraint 
\begin{equation}\label{E:radial}
  \left|R_{\hat{1}'\hat{0}'\hat{1}'\hat{0}'}\right|=
  \left|R_{\hat{r}\hat{t}\hat{r}\hat{t}}\right|\\
    =e^{-2\alpha(r)}
      \left|\gamma''(r)-\alpha'(r)\gamma'(r)
      +\left[\gamma'(r)\right]^2\right|
   \le (10^8\,\text{m})^{-2}.
\end{equation}
The lateral tidal constraints are (reinserting $c$)
\begin{multline}\label{E:lateral}
  \left|R_{\hat{2}'\hat{0}'\hat{2}'\hat{0}'}\right|
  =\left|R_{\hat{3}'\hat{0}'\hat{3}'\hat{0}'}\right|
  =\gamma^2\left|R_{\hat{\theta}\hat{t}\hat{\theta}\hat{t}}\right|
  +\gamma^2\left(\frac{v}{c}\right)^2\left|
     R_{\hat{\theta}\hat{r}\hat{\theta}\hat{r}}\right|\\
   =\gamma^2\left(\frac{1}{r}e^{-2\alpha(r)}\gamma'(r)\right)
   +\gamma^2\left(\frac{v}{c}\right)^2
       \left(\frac{1}{r}e^{-2\alpha(r)}\alpha'(r)\right)
     \le (10^8\,\text{m})^{-2};
\end{multline} 
here $\gamma^2=1/\left[1-(v/c)^2\right]$.

As already noted, wormhole solutions allowed by general relativity 
may be subject to severe constraints from quantum field theory.  Of
particular interest to us is Eq. (95) in Ref.~\cite{FR96} (returning 
to geometrized units):
\begin{equation}\label{E:QI}
  \frac{r_m}{r_0}\le\left(\frac{1}{v^2-b_0'}\right)^{1/4}
       \frac{\sqrt{\gamma}}{f}\left(\frac{l_p}{r_0}\right)^{1/2},
\end{equation}
where $r_m$ is the smallest of several length scales, $v$ is the 
velocity of a boosted observer relative to the static frame, $\gamma=
1/\sqrt{1-v^2}$, $l_p$ is the Planck length, $f$ is a small
scale factor, and $b_0'=b'(r_0)$.  Inequality (\ref{E:QI}) is 
trivially satisfied whenever $b'(r_0)\approx 1.$ Accordingly, we 
will impose this condition to help satisfy the tidal constraints 
and to obtain some of the numerical estimates.  The fact remains, 
however, that according to Ref.~\cite{FR05a}, since the WEC 
is violated arbitrarily close to the throat, the analysis 
should be extended, but to do so would require additional 
information about the redshift and shape functions.  We will 
therefore return to this problem in Sec.~\ref{S:QI}. 

Returning to Eq.~(\ref{E:shape}), 
we have for the shape function,
\begin{equation}\label{E:bprime}
    b'(r_0)=\frac{d}{dr}\left[r(1-e^{-2\alpha(r)})\right]_{r=r_0}\\
    =2r_0e^{-2\alpha(r_0)}\alpha'(r_0)+1-e^{-2\alpha(r_0)}.
\end{equation}
To obtain $b'(r_0)=1,$ we require that 
\[
   \lim_{r \to r_0}e^{-2\alpha(r)}\alpha'(r)=0.
\]
As a consequence, the radial tidal constraint (\ref{E:radial}) 
is satisfied at the throat, while the lateral tidal constraints 
(\ref{E:lateral}) merely constrain the velocity of the traveler
in the vicinity of the throat.

The condition $b'(r_0)\approx 1$ has two other consequences: since 
the inequality (\ref{E:QI}) is satisfied, there is no particular 
restriction on the size $r_0$ of the radius of the thoat.  On 
the other hand, $b'(r_0)\approx 1$ implies that the wormhole will 
flare out very slowly, so that the coordinate distance from 
$r=r_0$ to $r=r_1$ will be much less than the proper distance.  
(This behavior can be seen from Fig.~2.)
\section{The exotic region}
We saw in the last section that $\alpha$ has to go to infinity fast 
enough so that $\lim_{r \to r_0}e^{-2\alpha(r)}\alpha'(r)=0.$  At 
the same time, $\alpha$ has to go to infinity slowly enough so that 
the proper distance  
\begin{equation*}
   \ell(r)=\int\nolimits_{r_0}^{r}e^{\alpha(r')}dr'
\end{equation*}
is finite.  Then by the mean-value theorem, there exists a value 
$r=r_2$ such that
\[
   \ell(r)=e^{\alpha(r_2)}(r-r_0),\quad r_0<r_2<r.
\]
In particular, $\ell(r_0)=0$ and 
\begin{equation}\label{E:meanvalue}
  \ell(r_1)=e^{\alpha(r_2)}(r_1-r_0).
\end{equation}
With this information we can examine the radial tidal constraint 
at $r=r_1.$  From Eq.~(\ref{E:Riemann1})
\begin{multline*}\label{E:radial1}
  |R_{\hat{r}\hat{t}\hat{r}\hat{t}}|=e^{-2\alpha(r_1)}
   \left|\gamma''(r_1)-\alpha'(r_1)\gamma'(r_1)
      +\left[\gamma'(r_1)\right]^2\right|\\
     =e^{-2\alpha(r_1)}\left|\gamma''(r_1)
       -\alpha'(r_1)[-\alpha'(r_1)]
         +\left[\alpha'(r_1)\right]^2\right|
\end{multline*}
by Eq.~(\ref{E:FR1}).  So by inequality (\ref{E:radial}), 
\begin{equation*}
 |R_{\hat{r}\hat{t}\hat{r}\hat{t}}|
    =e^{-2\alpha(r_1)}\left|\gamma''(r_1)+
     \alpha'(r_1)\alpha'(r_1)
         +\left[\alpha'(r_1)\right]^2\right|\\
            \le (10^8\text{m})^{-2}.
\end{equation*}
Since $e^{-2\alpha(r)}$ is strictly increasing, it follows that
\[
     e^{-2\alpha(r_2)}\left|\gamma''(r_1)
       +2\left[\alpha'(r_1)\right]^2\right|
            <10^{-16} \text{m}^{-2}.
\]
From Eq.~(\ref{E:meanvalue}), we now get the following:
\[
   \frac{(r_1-r_0)^2}{[\ell(r_1)]^2}
    \left|\gamma''(r_1)+2\left[\alpha'(r_1)\right]^2\right|
            <10^{-16}\text{m}^{-2}
\]
and
\begin{equation}\label{E:abs1}
  \left|\gamma''(r_1)+2\left[\alpha'(r_1)\right]^2\right|
  <\frac{[\ell(r_1)]^2}{10^{16}(r_1-r_0)^2}.
\end{equation}
As a consequence,
\begin{equation}\label{E:abs2}
  \gamma''(r_1)+2\left[\alpha'(r_1)\right]^2
   <\frac{[\ell(r_1)]^2}{10^{16}(r_1-r_0)^2}
\end{equation}
or
\begin{equation}\label{E:abs3}
  \gamma''(r_1)+2\left[\alpha'(r_1)\right]^2
   >-\frac{[\ell(r_1)]^2}{10^{16}(r_1-r_0)^2}.
\end{equation}
So if either condition (\ref{E:abs2}) or condition (\ref{E:abs3})
is satisfied, then so is condition (\ref{E:abs1}).

To estimate the size of the exotic region, we need some 
idea of the magnitude of $\alpha(r_1)$ and $\alpha'(r_1)$.  
The only information available is that $\alpha(r)$ increases 
slowly enough to keep 
$\int\nolimits_{r_0}^{r}e^{\alpha(r')}dr'$ finite.  One way to 
accomplish this is to assume that for computational purposes 
$\alpha(r)$ is roughly logarithmic, at least near $r=r_1$, as 
in Ref.~\cite{pK06}: 
\begin{equation}\label{E:log}
   \alpha(r)\sim \text{ln}\frac{K}{(r-r_0)^A},\quad 
             0<A<1.
\end{equation}

While this form of $\alpha(r)$ may be just a computational 
convenience, there is no guarantee that $\alpha(r)$ is 
drastically different from the ln-form, so that one must 
proceed with caution.  In particular, we now have to assume 
that
\begin{equation}\label{E:D2alpha}
    \alpha'(r_1)\sim -\frac{A}{r_1-r_0}\quad \text{and}\quad
     \alpha''(r_1)\sim \frac{A}{(r_1-r_0)^2}.
\end{equation}
Since we also want $\alpha''(r)>|\gamma''(r)|$ [or
$\alpha''(r)>-\gamma''(r)$], we have in view of 
inequality (\ref{E:abs2}),  
\begin{equation}\label{E:main}
  \frac{A}{(r_1-r_0)^2}>-\gamma''(r_1)>\frac{2A^2}{(r_1-r_0)^2}-
     \frac{[\ell(r_1)]^2}{10^{16}(r_1-r_0)^2}.  
\end{equation}
Conversely, the inequality 
\begin{equation*}
   \frac{2A^2}{(r_1-r_0)^2}-\frac{[\ell(r_1)]^2}
          {10^{16}(r_1-r_0)^2}\\
    =2[\alpha'(r_1)]^2-\frac{[\ell(r_1)]^2}{10^{16}(r_1-r_0)^2}
    <-\gamma''(r_1)  
\end{equation*}
implies condition (\ref{E:abs2}).  Since $-\gamma''(r)<
\alpha''(r)$, we conclude that inequality (\ref{E:main}) 
is valid if, and only if, condition (\ref{E:abs2}) is met.

Inequality (\ref{E:main}) now implies that 
\begin{equation}
   2A^2-A-\frac{[\ell(r_1)]^2}{10^{16}}<0.
\end{equation}
Equality is achieved whenever
\[
    A=\frac{1\pm\sqrt{1+\frac{8[\ell(r_1)]^2}{10^{16}}}}{4}.
\]
Returning to the condition $b'(r_0)=1$ for a moment, if the 
ln-approximation is used again, then $A$ must exceed 1/2.  It 
follows that 
\begin{equation}\label{E:Afinal}
   \frac{1}{2}<A<\frac{1+\sqrt{1+\frac{8[\ell(r_1)]^2}{10^{16}}}}{4}.
\end{equation}
This solution shows that considerable fine-tuning is required.  We 
will return to this point in Sec.~\ref{S:finetune}.

Finally, observe that with the extra condition $|\gamma''(r_1)|
<\alpha''(r_1)$, the  qualitative features in Fig.~1 are retained,
so that no additional assumptions are needed.

Returning to Eq.~(\ref{E:log}), the form of $\alpha(r)$ yields the 
following estimate for $\ell(r)$:
\begin{equation}\label{E:properfinal}
   \ell(r)\sim\int^{r}_{r_0}e^{\,\text{ln}[K/(r'-r_0)^A]}dr'\\
    =\frac{K}{1-A}(r-r_0)^{1-A}, \quad 0<A<1.
\end{equation}
(See Fig.~2.)  When $A\approx 1/2$, then $\ell(r)\approx
2K(r-r_0)^{1/2}.$
We can see from the figure that $\ell(r_1)$ is much larger than 
$r_1-r_0$ near the throat, a consequence of the slow flaring out. 
\begin{figure}[htbp]
\begin{center}
\includegraphics[clip=true, draft=false, bb=0 0 299 212, angle=0, width=4.5in, height=3.0in, 
   viewport=50 50 296 200]{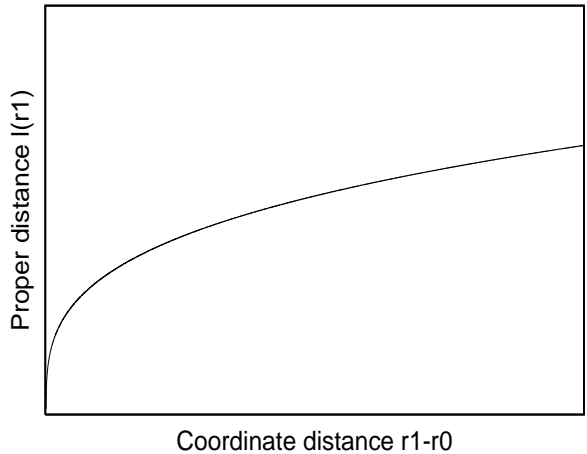}
\end{center}
\caption{\label{fig:figure2}Graph showing the proper thickness $\ell(r_1)$
     as a function of the coordinate distance $r_1-r_0$.}
\end{figure} 

\section{Numerical estimates}\label{S:numerical}
This section is devoted to numerical calculations.  As in 
Ref.~\cite{MT88}, we start with 
an estimate of the size of the wormhole, as measured by the 
position of the space station.  According to Ref.~\cite{MT88}, 
the space station should be far enough away from the throat so 
that 
\begin{equation}\label{E:flat}
   1-\frac{b(r)}{r}=e^{-2\alpha(r)}\approx 1,
\end{equation}
making the space nearly flat.  Another condition involves the 
redshift function: at the station we must also have
\begin{equation}\label{E:station1}
   |\gamma'(r)|\le g_{\oplus}/\left(c^2\sqrt{1-b(r)/r}\right).
\end{equation}
It will be seen below that for our wormhole the first condition,
Eq.~(\ref{E:flat}), is easily satisfied.  By condition 
(\ref{E:FR3}), as well as Fig. 1, $|\alpha'(r)|<\gamma'(r)$ for
$r>r_1$.  So if $1-b(r)/r\approx 1,$ then we have 
\begin{equation}\label{E:station2}
    |\alpha'(r)|<10^{-16}\,\text{m}^{-1}
\end{equation}  
at the station.  This inequality should give us at least a rough 
estimate of the distance to the station, since for large $r$,
$\alpha(r)\sim\gamma(r).$  It must be kept in mind, however, 
that the inequality $|\alpha'(r)|<\gamma'(r)$ implies that this 
procedure does underestimate the distance, perhaps by quite a 
bit.  The main reason for using $\alpha$ in the first place is 
to avoid making additional assumptions involving $\gamma$.  
Instead, $\gamma$ can be left to its more obvious role, 
adjusted if necessary, to help meet the tidal constraint in  
Eq.~(\ref{E:radial}) for $r>r_1$.  We will return to this point 
after discussing $\alpha.$  (Very close to the throat the 
redshift function may also have to be fine-tuned to help meet 
the quantum inequalities, as will be seen in Sec.~\ref{S:QI}.)

To make use of condition (\ref{E:station2}),   we will again assume  
that $\alpha(r)$ is similar to $\text{ln}[K/(r-r_0)^A]$ 
near $r=r_1$, to be denoted by $\alpha_{\text{left}}(r)$.   
To ensure asymptotic flatness, this function will be joined 
smoothly at some $r=r_3>r_1$ to a function 
$\alpha_{\text{right}}(r)$, assumed to have the form 
$\alpha_{\text{right}}(r)=C/(r-r_0)^n$.  Thus
\[
  \alpha_{\text{left}}(r_3)=\alpha_{\text{right}}(r_3)\quad 
  \text{and}  \quad \alpha'_{\text{left}}(r_3)=
           \alpha'_{\text{right}}(r_3).
\]
From
\[
  \alpha'_{\text{left}}(r_3)=-\frac{A}{r_3-r_0}=-\frac{nC}
     {(r_3-r_0)^{n+1}}=\alpha'_{\text{right}}(r_3)
\]
we obtain $C=(A/n)(r_3-r_0)^n.$  Thus
\begin{equation}\label{E:right}
  \alpha_{\text{right}}(r)=\frac{(A/n)(r_3-r_0)^n}
       {(r-r_0)^n}.
\end{equation}
From $\alpha_{\text{left}}(r_3)=\alpha_{\text{right}}(r_3),$ 
we get 
\[
  e^{\text{ln}[K/(r_3-r_0)^A]}=
  e^{[(A/n)(r_3-r_0)^n]/(r_3-r_0)^n}=e^{A/n}
\]
or
\begin{equation}\label{E:K}
   K=e^{A/n}(r_3-r_0)^A.
\end{equation}
The desired distance $r=r_s$ to the space station can now be 
estimated using Eq.~(\ref{E:right}):
\[
  \left|\alpha'_{\text{right}}(r_s)\right|=
   \frac{A(r_3-r_0)^n}{(r_s-r_0)^{n+1}}=10^{-16}\,
    \text{m}^{-1},
\]
which implies that (since $r_s-r_0\approx r_s$)
\begin{equation}\label{E:station}
   r_s\approx \left[ 10^{16}A(r_3-r_0)^n\right]^{1/(n+1)}.
\end{equation}
For convenience we now restate Eq.~(\ref{E:properfinal}) 
for $r=r_1$:
\begin{equation}\label{E:restate}
   \ell(r_1)\approx \frac{K}{1-A}(r_1-r_0)^{1-A},
\end{equation}
where $K=e^{A/n}(r_3-r_0)^A.$

Given the resulting infinite number of solutions, how should 
the various parameters be chosen?  We know that the wormhole 
flares out very slowly at the throat, which suggests assigning 
a small coordinate distance to the exotic region, at least 
initially.  A good choice is $r_1-r_0=0.000001\,\text{m},$ 
as in Ref.~\cite{pK06}.  The distance $r_3-r_0$ can be much 
larger; so to fix ideas, we arbitrarily choose $r_3-r_0=
1\,\text{mm}=0.001\,\text{m}.$  Using Eqs. (\ref{E:station}) 
and (\ref{E:restate}) with $A=1/2$ for the calculations, the 
values of $\ell(r_1)$ and $r_s$ for various choices of $n$ are 
given in the accompanying table.     
\begin{table}
\begin{center}
\begin{tabular}{c|c|}
\phantom{n=0.60}&A=0.50\\ \hline
n=0.60&0.0146 cm\\
\phantom{n=0.60}&490 000 km\\ \hline
n=0.65&0.0136 cm\\
\phantom{n=0.60}&215 000 km\\ \hline
n=0.70&0.0129 cm\\
\phantom{n=0.60}&100 000 km\\ \hline
n=0.75&0.0123 cm\\
\phantom{0.06}&48 000 km\\ \hline
n=0.80&0.0118 cm\\
\phantom{n=0.60}&24 000 km\\ \hline
n=0.85&0.0114 cm\\
\phantom{n=0.60}&13 000 km\\ \hline

\end{tabular}
\caption{The top and bottom values in each cell are $\ell(r_1)$
and $r_s$, respectively.}
\end{center}
\end{table}

While our choices are necessarily somewhat arbitrary and the 
calculated values only approximate, the table allows a 
conservative estimate for both the proper thickness of the 
exotic region and the size of the wormhole, as measured by 
$r_s$, the distance to the space station.  Judging from the 
middle of the table, the proper thickness of the exotic 
region is only about 0.1 mm; $r_s$ need not be more than 
about 100 000 km --- and could even be much less.

Returning to the radial tidal constraint, based on experience 
with specific functions (as in Ref.~\cite{pK02}), 
$\left|R_{\hat{r}\hat{t}\hat{r}\hat{t}}\right|$ is likely to 
reach its peak just to the right of $r=r_1$.  The simplest 
way to handle this problem is to 
tighten the constraint in Eq.~(\ref{E:radial}) at $r=r_1$ by 
reducing the right side.  This change increases the degree of 
fine-tuning in condition (\ref{E:Afinal}).  

A final consideration is the time dilation near the throat. 
Denoting the proper distance by $\ell$ and the proper time by 
$\tau$, as usual, we let $v=d\ell/d\tau$, so that 
$d\tau=d\ell/v$, assuming now that $\gamma\approx 1$.  
Since $d\ell=e^{\alpha(r)}dr$ and $d\tau=e^{\gamma(r)}dt$, 
we have for any coordinate interval $\Delta t$:
\begin{equation*}
  \Delta t=\int\nolimits_{t_a}^{t_b}dt=
     \int\nolimits_{\ell_a}^{\ell_b}e^{-\gamma(r)}\frac{d\ell}{v}= 
     \int\nolimits_{r_a}^{r_b}\frac{1}{v}e^{-\gamma(r)}e^{\alpha(r)}
        dr.     
\end{equation*}
If we assume, once again, that 
$\alpha(r)\sim\text{ln}[K/(r-r_0)^A]$, then on the interval 
$[r_0,r_1]$, 
\begin{equation*}
   \Delta t= \int_{r_0}^{r_1}\frac{1}{v}
    e^{-\gamma(r)}\frac{K}{(r-r_0)^{A}}dr.
\end{equation*}
Since $\gamma(r)$ is finite, the small size of the interval 
$[r_0,r_1]$ implies that $\Delta t$ is relatively small.

\section{Additional models}\label{S:additional}
We assumed in the previous section that for computational 
purposes, $\alpha(r)$ is roughly logarithmic.  In this section 
we consider a more complicated class of functions for $\alpha$:
\begin{equation}\label{E:hyper}
  \alpha(r)=a\,\text{ln}\left(\frac{1}{(r-r_0)^b}
    +\sqrt{\frac{1}{(r-r_0)^{nb}}+1}\right).
\end{equation}
The main advantage of this model is that $\alpha(r)\rightarrow 0$ 
as $r\rightarrow \infty$, so that no modification is needed.  
For now we will concentrate on the special case $n=2$ and return 
to Eq.~(\ref{E:hyper}) later.  For $n=2$, the equation becomes
\[
    \alpha(r)=a\,\text{sinh}^{-1}\frac{1}{(r-r_0)^b},
        \quad b>\frac{1}{2a}.
\]
The need for the assumption $b>1/(2a)$ comes from the shape 
function 
\[
  b(r)=r\left(1-e^{-2a\,\text{sinh}^{-1}
   [1/(r-r_0)^b]}\right):
\]
\begin{multline*}
  b'(r)=1-e^{-2a\,\text{sinh}^{-1}[1/(r-r_0)^b]}\\
   +r\left(-e^{-2a\,\text{sinh}^{-1}[1/(r-r_0)^b]}\right)
    \frac{2ab}{(r-r_0)\sqrt{(r-r_0)^{2b}+1}};
\end{multline*}
$b'(r)\rightarrow 1$ as $r\rightarrow r_0$, as long as 
$b>1/(2a).$  To see this, it is sufficient to examine 
\[
  e^{-2a\,\text{sinh}^{-1}[1/(r-r_0)^b]}\frac{1}{r-r_0}
\]
as $r\rightarrow r_0$:
\begin{multline*}
  \frac{1}{\left[\frac{1}{(r-r_0)^b}+\sqrt{\frac{1}{(r-r_0)^{2b}}+1}
    \right]^{2a}}\frac{1}{r-r_0}\\
   =\frac{1}{\frac{1}{(r-r_0)^{2ab}}\left[1+(r-r_0)^b
    \sqrt{\frac{1}{(r-r_0)^{2b}}+1}\right]^{2a}}\frac{1}{r-r_0}\\
    =\frac{1}{\frac{1}{(r-r_0)^{2ab-1}}
    \left[1+\sqrt{1+(r-r_0)^{2b}}\right]^{2a}}\\
    \sim(r-r_0)^{2ab-1}, \,\,\text{whence}\,2ab-1>0.
\end{multline*}
For computational purposes, however, we will simply let $b=1/(2a)$.  
Consider next,
\[
  \alpha'(r)=-\frac{ab}{(r-r_0)\sqrt{(r-r_0)^{2b}+1}},\quad r>r_0,
\]
and
\[
   \alpha''(r)=\frac{ab\left[(1+b)(r-r_0)^{2b}+1\right]}
      {(r-r_0)^2\left[(r-r_0)^{2b}+1\right]^{3/2}}.
\]
Given that $r_1-r_0=0.000001\,\text{m}$ from Sec.~
\ref{S:numerical}, we get 
\[
    \alpha'(r_1)\approx -\frac{ab}{r_1-r_0}
\]
and 
\[
    \alpha''(r_1)\approx\frac{ab}{(r_1-r_0)^2}.
\]
Comparing these results to Eq.~(\ref{E:D2alpha}), we 
conclude, in view of inequality (\ref{E:abs2}) and 
$|\gamma''(r)|<\alpha''(r)$, that $ab$ is subject to exactly 
the same fine-tuning as $A$ in inequality (\ref{E:Afinal}):
\begin{equation}\label{E:abfinal}
  \frac{1}{2}<ab<\frac{1+\sqrt{1+\frac{8[\ell(r_1)]^2}{10^{16}}}}{4}.   
\end{equation}
The left inequality confirms that $b>1/(2a)$.

Letting $b=1/(2a)$, we now have 
\[
   \ell(r_1)=\int_{r_0}^{r_0+0.000001}
   e^{a\,\text{sinh}^{-1}[1/(r-r_0)^{1/(2a)]}}dr.
\]
These values change very little with $a$.  For example, if $a$ 
ranges from 0.1 to 0.5, then $\ell(r_1)$ ranges from 0.0021 m 
to 0.0028 m.  These values are larger than our previous values, 
unless we reduce the coordinate distance.  Thus for $r_1-r_0=
0.000000001\,\text{m}$ and $a=0.5$, we get $\ell(r_1)=
0.000089\,\text{m}<0.1\,\text{mm}$, corresponding to $r_s
\approx\,$70 000 km .

A good alternative is to use Eq.~(\ref{E:hyper}), subject to 
the condition
\[
    nab-b+\frac{1}{2}nb>1.
\]
(As before, this condition comes from the requirement that 
$b'(r_0)=1$; in fact, if $n=2,$ we are back to $2ab>1$.)  
For example, retaining $r_1-r_0=0.000001\,\text{m}$, if 
$a=0.2$ and $b=1$, then $nb=2.857$.  These values yield 
$\ell(r_1)\approx 0.0000725\,\text{m}<0.1\,\text{mm}.$  
The corresponding distance $r_s$, obtained from 
$\alpha'(r)$ [now referring to Eq.~(\ref{E:hyper})], 
is about 45 000 km.
 
Using the equation $nab-b+\frac{1}{2}nb=1$ to eliminate $n$ 
in Eq.~(\ref{E:hyper}) shows that further reductions in 
$\ell(r_1)$ are only significant if $a$ and $b$ get 
unrealistically close to zero.  So practically speaking, a  
further reduction in the proper distance $\ell(r_1)$ 
requires a reduction in the coordinate distance $r_1-r_0$.

\section{The fine-tuning problem in general}\label{S:finetune}
The almost identical inequalities (\ref{E:Afinal}) and 
(\ref{E:abfinal}) suggest that the degree of fine-tuning 
encountered is a general property of the type of wormhole 
being considered, namely wormholes for which $b'(r_0)=1$ and 
$\alpha(r)=a\,\text{ln}f(r-r_0),$ where (generalizing from 
earlier cases) $f(r-r_0)|_{r=r_0}$ is undefined $(+\infty)$ 
and $f(\frac{1}{r-r_0})|_{r=r_0}$ is a constant (possible 
zero).  If we also assume that $g(r-r_0)=f(\frac{1}{r-r_0})$
can be expanded in a Maclaurin series, then we have for 
$r\approx r_0$, 
\begin{multline*}
  f\left(\frac{1}{r-r_0}\right)=g(r-r_0)=a_0+a_1(r-r_0)\\
      +a_2(r-r_0)^2+\cdot\cdot\cdot
     \approx a_0+a_1(r-r_0).
\end{multline*}  
It follows that 
\[
    f(r-r_0)=a_0+\frac{a_1}{r-r_0}
\]
near the throat.  So
\[
  \alpha(r)=a\,\text{ln}\left(a_0+\frac{a_1}{r-r_0}\right),
\]
\begin{equation}\label{E:der1}
   \alpha'(r_1)=\frac{-aa_1}{a_0+\frac{a_1}{r_1-r_0}}
     \frac{1}{(r_1-r_0)^2}\sim -\frac{a}{r_1-r_0},
\end{equation}
and
\begin{equation}\label{E:der2}
   \alpha''(r_1)=\frac{aa_1[2a_0(r_1-r_0)+a_1]}
    {[a_0(r_1-r_0)^2+a_1(r_1-r_0)]^2}\\
      \sim\frac{a}{(r_1-r_0)^2}.
\end{equation}

To show that $b'(r_0)=1,$ we need to show that 
$e^{-2\alpha(r)}\alpha'(r)\rightarrow 0$ as $r\rightarrow 
r_0$:
\begin{multline*}
   e^{-2a\,\text{ln}[a_0+a_1/(r-r_0)]}\frac{-aa_1}
      {a_0+\frac{a_1}{r-r_0}}\frac{1}{(r-r_0)^2}\\
    =\frac{1}{\left(a_0+\frac{a_1}{r-r_0}\right)^{2a}}
     \frac{-aa_1}{a_0+\frac{a_1}{r-r_0}}
        \frac{1}{(r-r_0)^2}.
\end{multline*} 
Since $a_0$ is negligible if $r$ is close to $r_0$, we obtain 
\[
    e^{-2\alpha(r)}\alpha'(r)\sim(r-r_0)^{2a-1},
\]
so that $2a-1>0$ and $a>\frac{1}{2}.$  
Comparing Eqs. (\ref{E:der1}) and (\ref{E:der2}) to 
Eq.~(\ref{E:D2alpha}), we conclude that
\begin{equation}\label{E:finetuning}
    \frac{1}{2}<a<\frac{1+\sqrt{1+\frac{8[\ell(r_1)]^2}
         {10^{16}}}}{4}.
\end{equation}
So the amount of fine-tuning required really does appear to 
be a general property of wormholes of the present type.  While 
the degree of fine-tuning considered so far is quite severe, 
it is considerably milder than most of the cases discussed in 
Ref.~\cite{FR05a}.

As indicated at the end of Sec.~\ref{S:additional}, if 
Eq.~(\ref{E:hyper}) is used in the model, any further reduction 
in $\ell(r_1)$ requires a reduction in the coordinate distance 
$r_1-r_0$.  We can see from condition (\ref{E:finetuning}), 
however, that reducing $\ell(r_1)$ will increase the degree of 
fine-tuning.  While basically presenting us with an engineering 
challenge, this increase can only be carried so far.  In 
particular, we are confirming the assertion in 
Ref.~\cite{FR05a} that the amount of exotic matter cannot be made 
arbitrarily small.

\section{The quantum inequalities near the throat}\label{S:QI}
We know that inequality (\ref{E:QI}) is trivially satisfied as 
long as $b'(r_0)\approx 1.$  The purpose of this section is to 
derive an analogous inequality for $r$ close to $r_0$.

The analysis in Ref.~\cite{FR96} is based on the inequality 
\begin{equation}
   \frac{\tau_0}{\pi}\int^{\infty}_{-\infty}
   \frac{\langle T_{\mu\nu}u^{\mu}u^{\nu}\rangle d\tau}
    {\tau^2+\tau_0^2}\ge -\frac{3}{32\pi^2\tau_0^4},
\end{equation}
where $\tau$ is the observer's proper time and $\tau_0$ the 
duration of the sampling time.  (See Ref.~\cite{FR96} for 
details.)  Put another way, the energy density is sampled in a 
time interval of duration $\tau_0$ which is centered around an 
arbitrary point on the observer's worldline so chosen that 
$\tau=0$ at this point.  The sampling time itself is usually 
assumed to be so short that the energy density does not change 
very much over this time interval and may therefore be taken 
to be approximately constant:
\begin{multline}
   \frac{\tau_0}{\pi}\int^{\infty}_{-\infty}
   \frac{\langle T_{\mu\nu}u^{\mu}u^{\nu}\rangle d\tau}
    {\tau^2+\tau_0^2}\approx 
    \langle T_{\mu\nu}u^{\mu}u^{\nu}\rangle 
    \frac{\tau_0}{\pi}\int^{\infty}_{-\infty}
     \frac{d\tau}{\tau^2+\tau_0^2}\\
   =\langle T_{\mu\nu}u^{\mu}u^{\nu}\rangle=\rho 
     \ge -\frac{3}{32\pi^2\tau_0^4}.
\end{multline}    

To obtain a condition analogous to inequality (\ref{E:QI}), 
it is convenient to use the forms for $T_{\hat{t}\hat{t}}$
and $T_{\hat{r}\hat{r}}$ in Ref.~\cite{FR96}:
\begin{equation}
    T_{\hat{t}\hat{t}}=\rho=\frac{b'(r)}{8\pi r^2}
\end{equation} 
and
\begin{equation}
   T_{\hat{r}\hat{r}}=-\tau=p_r
   =-\frac{1}{8\pi}\left[\frac{b(r)}{r^3}-\frac{2\gamma'(r)}{r}
    \left(1-\frac{b(r)}{r}\right)\right].
\end{equation}
Away from the throat we have the analogous formula for $r_m$:
\begin{equation}
   r_m\equiv\text{min}\left[b(r), \left|\frac{b(r)}{b'(r)}\right|,
    \frac{1}{|\gamma'(r)|}, 
        \left|\frac{\gamma'(r)}{\gamma''(r)}\right|\right].
\end{equation} 
Finally, still following Ref.~\cite{FR96}, 
the energy density in the ``boosted frame" is by the 
Lorentz transformation 
\begin{equation}
   T_{\hat{0}'\hat{0}'}=\rho'=\gamma^2(\rho+v^2p_r),
\end{equation}
where $\gamma=1/\sqrt{1-v^2}.$
(It is stated in Ref.~\cite{FR96} that in this frame the energy 
density does not change very much over the sampling time,
 so that $\rho'\ge -3/(32\pi^2\tau_0^4)$.)
Substitution yields
\[
   \rho'=\frac{\gamma^2}{8\pi r^2}\left[b'(r)
   -v^2\frac{b(r)}{r}+v^2r(2\gamma'(r))
    \left(1-\frac{b(r)}{r}\right)\right].
\]
From
\[
    \frac{3}{32\pi^2\tau_0^4}\ge -\rho'
\]
we now have
\begin{equation*}
  \frac{32\pi^2\tau_0^4}{3}\le\\\frac{8\pi r^2}{\gamma^2}
   \left[v^2\frac{b(r)}{r}-b'(r)-v^2r(2\gamma'(r))
    \left(1-\frac{b(r)}{r}\right)\right]^{-1}.
\end{equation*}
The suggested sampling time is 
\[
   \tau_0=\frac{fr_m}{\gamma},
\]
where $f$ is a scale factor such that $f\ll 1$.  After dividing 
both sides by $r^4$, we have (disregarding a small coefficient)
\begin{equation*}
    \frac{f^4r_m^4}{r^4\gamma^4} \le\\\frac{1}{r^2\gamma^2}
    \left[v^2\frac{b(r)}{r}-b'(r)-2v^2r\gamma'(r)
        \left(1-\frac{b(r)}{r}\right)\right]^{-1}
\end{equation*}
and, after inserting $l_p$, 
\begin{equation}\label{E:genQI}
   \frac{r_m}{r}\le 
    \left(\frac{1}{v^2\frac{b(r)}{r}-b'(r)-2v^2\gamma'(r)
    \frac{r}{l_p}\left(1-\frac{b(r)}{r}\right)}\right)^{1/4}\\
     \times\frac{\sqrt{\gamma}}{f}
        \left(\frac{l_p}{r}\right)^{1/2}.
\end{equation}
At the throat, where $b(r_0)=r_0$, inequality (\ref{E:genQI}) 
reduces to inequality (\ref{E:QI}).

It is understood that in inequality (\ref{E:genQI}), $r$ is 
close to $r_0$, so that $b'(r)$ is close to unity.  But since 
$b'(r)<1$, inequality (\ref{E:QI}) is not necessarily satisfied 
for $r>r_0$.  Inequality (\ref{E:genQI}), however, $\gamma'(r)$ 
can be fine-tuned so that the condition is satisfied in the 
interval $[r_0,r_1]$.
\section{Conclusion}
This paper discusses a class of wormhole geometries that, 
finally, satisfy the constraints from quantum field theory, 
while striking a balance between reducing the proper thickness 
of the exotic region as much as possible, while trying to keep 
the fine-tuning requirement within reasonable bounds.  The 
assumptions on the metric coefficients in line element 
(\ref{E:line}) have been kept to a minimum.  An unexpected 
finding is that the degree of fine-tuning is a generic 
property of the type of wormhole discussed.

The wormholes are macroscopic and satisfy various 
traversability criteria.

There are many possible choices for the parameters and hence 
many solutions.  The particular choices discussed are fairly 
conservative, leading to the following promising results: 
approximately 0.1 mm for the proper thickness of the exotic 
region, corresponding roughly to a distance of 100 000 km to 
the space station, possibly much less.  The proper thickness 
of 0.1 mm should not be viewed as the final outcome, however.  
By decreasing the coordinate distance, it is theoretically 
possible to decrease the thickness of the exotic region 
indefinitely.  While this decrease may be thought of as an 
engineering problem, the fact remains that the concomitant 
increase in the degree of fine-tuning would eventually exceed 
any practical limit.

\end{document}